\documentclass[aps,amsmath,amssymb,preprintnumbers,groupedaddress,twocolumn]{revtex4}
\usepackage{graphicx}

\newcommand{\Z}{\mathbb{Z}}

\newcommand{\nc}{\newcommand}
\nc{\beq}{\begin{equation}}
\nc{\eeq}{\end{equation}}
\nc{\bea}{\begin{eqnarray}}
\nc{\eea}{\end{eqnarray}}

\def\gsim{\mathrel{\rlap{\lower4pt\hbox{\hskip1pt$\sim$}}
    \raise1pt\hbox{$>$}}}       

\def\K3{{\bf K3}}

\def\ov{\overline}

\def\n2d{\cN_{V^*}^{\otimes 2}}

\def\cN{{\mathcal N}}

\begin{document}

\preprint{MPP-207-93}
\preprint{UPR-1084-T}
\preprint{LMU-ASC 46/07}
\title{Non-perturbative Yukawa Couplings from String Instantons}

\author{Ralph Blumenhagen$^{1         }$} \email{blumenha@mppmu.mpg.de}
\author{Mirjam Cveti{\v c$^{2         }$}} \email{cvetic@cvetic.hep.upenn.edu}
\author{Dieter L\"ust$^{1,3         }$} \email{luest@mppmu.mpg.de}
\author{Robert Richter$^{2         }$}\email{rrichter@physics.upenn.edu}
\author{Timo Weigand$^{2         }$}\email{timo@physics.upenn.edu}
\affiliation{$^{1         }$Max-Planck-Institut f\"ur Physik, F\"ohringer Ring 6, 80805 M\"unchen, Germany\\
$^{2         }$Department of Physics and Astronomy,
University of Pennsylvania, Philadelphia, USA \\
$^{3         }$ Arnold-Sommerfeld-Center for Theoretical Physics, Department f\"ur Physik, Ludwig-Maximilians-Universit\"at M\"unchen, Theresienstrasse 37, 80333 M\"unchen, Germany}

\begin{abstract}
\noindent
Non-perturbative D-brane instantons can generate
perturbatively absent though phenomenologically relevant
couplings for Type II orientifold compactifications with D-branes.
We discuss the generation of the perturbatively vanishing
$SU(5)$ GUT Yukawa coupling of type
$\langle{\bf 10}\, {\bf 10}\, {\bf 5}_H\rangle $.
Moreover, for a simple globally consistent intersecting D6-brane
model, we discuss the generation of mass terms
for matter fields. This can serve as a mechanism for decoupling exotic matter.

\end{abstract}


\maketitle

\date{today}

\bigskip

\section{Introduction}

Non-perturbative mass generation by gauge instantons is essential to explain the
pattern of meson masses in QCD.
Specifically, QCD instantons generate a non-perturbative, effective fermion interaction \cite{'t Hooft:1976fv}
that involves $N_f$ flavors of quark fields and breaks the perturbative
$U(1)_A$ axial symmetry to a discrete $Z_{2N_F}$ subgroup. In this way
the mass of the $\eta'$ meson gets generated   \cite{Witten:1978bc,Veneziano:1979ec}. In addition,
gauge instantons of the weak $SU(2)$ gauge group are responsible for
baryon number violating processes in the Standard Model of particle physics.

Following these observations, it would be very interesting if also
some of the quark and lepton masses  are of non-perturbative origin
in theories  beyond the Standard Model. For instance, S. Weinberg
\cite{Weinberg:1981mb} suggested that quark masses could be due to
gauge instantons of some hypercolour interactions in subquark
models. 

Also in string theory compactifications, quark and lepton masses, or rather
the respective
Yukawa couplings, can be generated by non-perturbative effects, more precisely by
string world-sheet \cite{Dine:1986zy} or by spacetime instantons. Most importantly, string theory
opens up some new perspectives in that non-perturbative effects do not only include
gauge instantons of the effective gauge theory, but also entirely stringy instantons
not related to effective gauge interactions.
In fact, during the last year there has been some progress towards a better
understanding of non-perturbative effects in ${\cal N}=1$ supersymmetric
four-dimensional string compactifications on Calabi-Yau
orientifolds
(\cite{Blumenhagen:2006xt,Haack:2006cy,Florea:2006si,Ibanez:2006da} and further developments in \cite{Abel:2006yk,Akerblom:2006hx,Cvetic:2007ku,Bianchi:2007fx,Argurio:2007vq,Bianchi:2007wy,Ibanez:2007rs,Akerblom:2007uc,Antusch:2007jd,recomb.}; for  closely related  earlier work see \cite{Ganor:1996pe,Billo:2002hm}).
Type IIA orientifolds with intersecting
D6-branes
(see \cite{Uranga:2003pz,Kiritsis:2003mc,Lust:2004ks,Blumenhagen:2005mu,Blumenhagen:2006ci,Marchesano:2007de} for reviews)
receive non-perturbative corrections from Euclidean D2-brane instantons,
short E2-instantons, wrapping
special Lagrangian three-cycles of the internal Calabi-Yau space.

Once a CFT description of the background is available, the induced non-perturbative couplings can be studied using methods from open string theory.
An analysis of the zero mode structure
of such instantons shows that so-called O(1)-instantons  \cite{Argurio:2007vq,Bianchi:2007wy,Ibanez:2007rs,recomb.}  can generate
terms in the effective four-dimensional superpotential.

Of particular interest are those induced interactions which are forbidden perturbatively
due to global $U(1)$ selection
rules.
Under suitable circumstances, E2-instantons can break these global $U(1)$ symmetries
to certain discrete subgroups and generate $U(1)$ violating interactions \cite{Blumenhagen:2006xt,Ibanez:2006da}.
One important example of these new couplings are non--perturbative Majorana mass
terms for right-handed neutrinos or $\mu$-terms in the MSSM Higgs sector \cite{Blumenhagen:2006xt,Ibanez:2006da,Cvetic:2007ku,Ibanez:2007rs,Antusch:2007jd}.

The existence of such perturbatively forbidden Type IIA couplings
resolves one of the puzzles about the proposed large coupling dual
description in terms of M-theory compactifications on singular
$G_2$-manifolds \cite{Atiyah:2001qf,Witten:2001uq}. In this picture, non-abelian gauge symmetries are
localised at an ADE-singularity over a supersymmetric three-cycle on
the $G_2$-manifold. Clearly, the perturbative $U(1)$ gauge
symmetries on the Type IIA D6-branes become massive due to the
Green-Schwarz mechanism and therefore decouple completely in the
strong coupling M-theory dual description, but how then should the
resulting global $U(1)$ selection rules appear? The resolution to
this puzzle  is given by the appearance of the described $U(1)$
breaking non-perturbative terms in the Type IIA picture. Therefore,
each coupling present in M-theory vacua on $G_2$-manifolds should be
realised either perturbatively or non-perturbatively in the Type IIA
orientifold.

In this paper we explore two new types of phenomenologically
important instanton generated couplings for Type II orientifolds. In
particular, we investigate the generation of the crucial
$SU(5)$-like GUT model Yukawa couplings of type ${\bf 10}\cdot{\bf
10}\cdot{\bf 5}_H$, which are known to be absent perturbatively
\cite{Blumenhagen:2001te}. In the second part of the letter we
provide a globally consistent intersecting D6-brane model and show
that an E2-instanton generates a mass term for certain matter
fields, thus providing a new mechanism for decoupling exotic matter.

\section{$SU(5)$ Yukawa couplings}

Grand Unified $SU(5)$-like models based on intersecting $D6$-branes
generically suffer from the absence of the  important Yukawa coupling
${\bf 10}\cdot{\bf 10}\cdot{\bf 5}_H$ and are therefore so far not
considered realistic. Such models were first generally proposed in
\cite{Antoniadis:2000en} and explicitly constructed for intersecting
D6-branes in
\cite{Blumenhagen:2001te,Ellis:2002ci,Cvetic:2002pj,Chen:2005ab,Cvetic:2006by,Gmeiner:2006vb}.

The minimal intersecting $D6$-brane model realizing $SU(5)$ GUT
is shown in table \ref{tablegut}.
\begin{table}[ht]
\centering
\begin{tabular}{|c|c|c|c|}
\hline
sector & number &  $U(5)_a\times U(1)_b$ reps. & $U(1)_X$   \\
\hline \hline
$(a',a)$ &  $3 + (1,1)$ & ${\bf 10}_{(2,0)}$   & ${1\over 2}$   \\
$(a,b)$ &  $3$ & $\ov{\bf 5}_{(-1,1)}$   & $-{3\over 2}$   \\
$(b',b)$ &  $3$ & ${\bf 1}_{(0,-2)}$ & ${5\over 2}$     \\
$(a',b)$ &  $1$ & ${\bf 5}^H_{(1,1)}+\ov{\bf 5}^H_{(-1,-1)}$ & $(-1) + (1)$  \\
\hline
\end{tabular}
\caption{GUT $SU(5)$ intersecting D6-brane model, $U(1)_X={1\over 4} U(1)_a -{5\over 4} U(1)_b$. The multiplet ${\bf 10}_{(2,0)}$ also contains the GUT Higgs field which should appear as a vector-like pair.
\label{tablegut} } 
\end{table}
Such a model involves only two stacks $a$ and $b$ of
branes giving rise to a $U(5)_a\times U(1)_b$ gauge symmetry.
The $U(5)_a$ splits into $SU(5)_a\times U(1)_a$, so that there
are two abelian gauge groups $U(1)_a\times U(1)_b$.
One linear combination of these is  anomalous
and becomes massive via the generalised Green-Schwarz mechanism.
However, it survives as a global symmetry in the effective action. 
Matter fields transforming
as $\bf{10}$ under $SU(5)_a$ arise at the intersections of stack $a$ with its
image $a'$, while the matter fields transforming as $\bf{\bar{5}}$ 
as well as Higgs
fields $\bf{5}_H$ and $\bf{\bar{5}}_H$ are located at intersections of stack
$a$ with $b$ and $b'$. 
For a globally consistent
model the concrete wrapping numbers decide if the anomaly free combination 
$U(1)_X$ of the  abelian groups really remains massless. 
If not, the model is of the usual Georgi-Glashow type, 
while in the presence of a massless $U(1)_X$ it represents a 
flipped $SU(5)$ model.

From the $U(1)_{a,b}$ charges it is clear that perturbatively
the two Yukawa couplings
\bea
 \langle{\bf 10}_{(2,0)}\, \ov{\bf 5}_{(-1,1)}\, \ov{\bf 5}^H_{(-1,-1)}\rangle,
  \quad\quad
   \langle \ov{\bf 5}_{(-1,1)}\, {\bf 1}_{(0,-2)}\,  {\bf 5}^H_{(1,1)}\rangle
\eea are present. Focussing for concreteness on flipped $SU(5)$,
these give masses to the heavy (u,c,t)-quarks and the leptons.
However, the Yukawa couplings for the light (d,s,b)-quarks \bea
\label{Yuk1}
 \langle{\bf 10}_{(2,0)}\, {\bf 10}_{(2,0)}\, {\bf 5}^H_{(1,1)}\rangle
\eea
are not invariant under the two $U(1)$s. Note that this
interaction is also of key importance for the solution of
 the doublet-triplet splitting problem
for flipped $SU(5)$. For a non-zero VEV of the Standard Model
singlet component in ${\bf 10}+\ov{\bf 10}$ there is
no partner for the weak Higgs doublet to pair up with.
For Georgi-Glashow $SU(5)$ models, the
role of (u,c,t) and (d,s,b)-quarks has to be interchanged
and the GUT Higgs field is usually in the adjoint representation
of $SU(5)$.

\begin{figure}[h]
\begin{center}
 \includegraphics[width=0.4\textwidth]{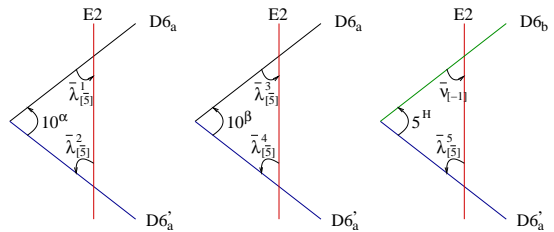}
\end{center}
\caption{\small Absorption of zero modes}\label{yukafig}
\end{figure}

Our main result is that the coupling (\ref{Yuk1}) can be
generated by an E2-instanton of suitable zero mode structure.
Concretely, the instanton has to wrap a rigid three-cycle $\Xi$
invariant under the orientifold projection $\Omega\ov\sigma$
 and carrying gauge group
$O(1)$ \cite{Argurio:2007vq,Bianchi:2007wy,Ibanez:2007rs,recomb.}. This
guarantees that the uncharged part of the instanton measure only
contains the four bosonic and two fermionic modes $ x^{\mu},
\theta^{\alpha}$ required for superpotential contributions. Now,
from the arguments in
\cite{Blumenhagen:2006xt,Florea:2006si,Ibanez:2006da} the coupling
(\ref{Yuk1}) requires in addition charged fermionic zero modes at
intersections between $\Xi$ and the D6-branes. These are
responsible for an effective $U(1)$ charge of the instanton which
can compensate for the excess of $U(1)$ charge of the operator
(\ref{Yuk1}). For intersection numbers \bea [\Xi\cap
\Pi_a]^+=[\Xi\cap \Pi_b]^+=0, \quad [\Xi\cap \Pi_a]^-=[\Xi\cap
\Pi_b]^-=1 \nonumber \eea we get five zero modes $\ov\lambda^i_{[\ov
5]}$ from the intersection of the instanton with $D6_a$ and one zero
mode $\ov\nu_{[-1]}$ from the intersection with $D6_b$. The
computation of the resulting couplings can be performed following
the prescription proposed in \cite{Blumenhagen:2006xt} and
exemplified for a concrete local model in \cite{Cvetic:2007ku}. Since the instanton lies in an
$\Omega\ov\sigma$ invariant position, one can absorb these six
matter zero modes with the three disc diagrams depicted in figure
\ref{yukafig}.
All charge selection rules are satisfied.  Each tree-level coupling
is by itself a sum over world-sheet instantons connecting the three
intersection points in the disc diagrams like
$10^\alpha_{[ij]}\,\ov\lambda^i\ov\lambda^j$, where $\alpha =1,2,3$
denotes the generation index. These discs induce open string
dependent terms in the instanton moduli  action of type 
\bea
\exp\left( -S_{mod}\right) = \exp\left( C^{10}_{\alpha}\,
10^\alpha_{[ij]}\,\ov\lambda^i\ov\lambda^j + C^{5}\,
5_{m}\,\ov\lambda^m \ov\nu \right) 
\eea 
which are integrated over the
charged fermionic measure $\int d^5 \ov \lambda \,d \ov \nu $. Due
to its Grassmannian nature, the index structure of the Yukawa
coupling is \bea \label{W_Y}
    W_Y=  Y^{\alpha\beta}_{\langle{\bf 10}\, {\bf 10}\, {\bf 5}_H\rangle} \,
     \epsilon_{ijklm}\,\, {\bf 10}^\alpha_{ij}\, {\bf 10}^\beta_{kl}\, {\bf 5}^H_{m} \,\, e^{- S_{E2}} \, \, e^{Z'} \, \,,
\eea
where the instanton action can be written as
$S_{E2}= {2\pi\over \alpha_{GUT}} { {\rm Vol}_{E2}\over {\rm Vol}_{D6} }$.
Here we have used that the volume of the $D6_a$-brane  determines the
gauge coupling at the GUT scale. Note that the ratio
${ {\rm Vol}_{E2}/ {\rm Vol}_{D6}}$ depends only on
the complex structure moduli, which are known to be constrained by the
D-term supersymmetry conditions for the D6-branes.
The superpotential coupling ${W_Y}$ also depends on the holomorphic part
of the one-loop determinant $ e^{Z'}$ arising from the annulus and M\"obius
diagrams ending on the  instanton and the D6-branes or O-plane, respectively \cite{Blumenhagen:2006xt}.
As shown in \cite{Abel:2006yk,Akerblom:2006hx,Akerblom:2007uc},
these are related to  one-loop gauge threshold corrections \cite{Lust:2003ky,Akerblom:2007np}.

One observes that the
replication of the zero modes $\ov \lambda_{i}$ is entirely due to
Chan-Paton indices so that each of the discs in figure \ref{yukafig}
depends only on the family index and not on the pair of zero modes
to which the open string operator couples.
Therefore the  final instanton generated Yukawa coupling factorises into
\bea
              Y^{\alpha\beta}_{\langle{\bf 10}\, {\bf 10}\, {\bf 5}_H\rangle}=
              Y^{\alpha} \, Y^{\beta}
  \eea
and the induced mass matrix for the quarks is always of unit rank.
In order to exhibit
non-perturbative masses for all three generations the model
therefore has to possess three independent E2-instanton sectors.

Concerning the suppression scale of the instanton
generated Yukawa coupling, for
$\alpha_{GUT}=1/24$ and ${ {\rm Vol}_{E2}/ {\rm Vol}_{D6}}=(R_{E2}/R_{D6})^3$
with the moderate suppression $R_{D6}={7\over 2}\, R_{E2}$,
the main instanton suppression factor is
$\exp\left(-S_{E2}\right) \simeq 3\cdot 10^{-2}$.
Since the E2-instanton lies in a $\ov\sigma$ invariant position\,
it seems natural that the length scale of the internal volume
is smaller than that of the $U(5)$ stack of D6-branes.

To summarise, we find that D-brane instantons can generate the
$\langle{\bf 10}\, {\bf 10}\, {\bf 5}_H\rangle$ Yukawa coupling. The
described mechanism works both for Georgi-Glashow as well as flipped
$SU(5)$ models. It is particularly attractive  for the case of
flipped SU(5): Here the E2-instanton not only generates the desired
couplings, but the complex structure dependent exponential
suppression $\exp(-S_{E2})$ can explain, as a bonus, the hierarchy
between the $(u,c,t)$ quarks and the $(d,s,b)$ quarks.

\section{Instanton generated mass terms}
\label{sec_model}

Most semi-realistic string models constructed so far come with exotic
vector-like states. For the phenomenological features of such models
it is  important to know whether these
states can become massive. To date mostly perturbative mechanism
have been discussed in the literature for generating such mass terms.
In this section we demonstrate for a concrete globally consistent
model that E2-instantons can also generate such mass terms.
We are working with  a Type IIA orientifold background
which serves as a simple model
based on $U(4)$ gauge symmetry  with a certain number of
matter fields in the  anti-symmetric representation of $U(4)$. For a similar global model in Type I theory see \cite{Bianchi:2007fx,Bianchi:2007wy}.

Concretely, we consider the orientifold $T^6/\Z_2\times \Z_2'$ with Hodge
numbers $(h_{11}, h_{12}) = (3,51)$. We employ the notation of
\cite{Blumenhagen:2005tn}, to which we refer for details of the
geometry and the construction of rigid cycles 
(see also \cite{Dudas:2005jx}). The orbifold group is
generated by $\theta$ and $\theta'$ acting as reflection in the
first and last two tori, respectively.

Table \ref{wrapping number} displays the wrapping numbers of the
simplest  globally consistent, supersymmetric model for the
choice that 
the $O6$-plane lying parallel to the instanton
is an $O6^+$-plane with the other three being $O6^-$-planes.

\begin{table}[htb]
\begin{center}
\begin{tabular}{|c||c|c|c|}
    \hline \rm{stack} & $N$ & $(n^1,m^1)\times (n^2,m^2)\times
(n^3,{m}^3)$ & $I_{E2x}$ \\
\hline \hline
    $U(4)$ &  8 & $(1,-1)\times (1,1)\times (1,1)$  &1 \\
\hline \hline
$E_2$ & 1 & $(1,0)\times (0,1)\times (0,-1)$ & \\
 \hline
\end{tabular}
\end{center}
\caption{ \label{wrapping number}
 Wrapping numbers of  $U(4)$ global
model.}
\end{table}

It involves only one stack of four bulk $D6$-branes (and its
orientifold image)
carrying $U(4)$ gauge group with three superfields in the adjoint
representation.
One can easily check that all consistency conditions are indeed
satisfied and supersymmetry fixes the complex structure moduli to the sublocus
$U^1-U^2-U^3=U^1\, U^2\, U^3 $.
The model has also  32  chiral superfields in the conjugate
anti-symmetric representation $\bf {\bar 6}$
of $U(4)$. Note that the  $\bf {6}$  of $SU(4)$ is
a real representation so that these states are chiral
only with respect to the diagonal $U(1)\subset U(4)$.
Since $U(1)$'s can be broken by instantons, there is
a chance that mass terms are non-perturbatively generated.

\begin{figure}[h]
\begin{center}
 \includegraphics[width=0.28\textwidth]{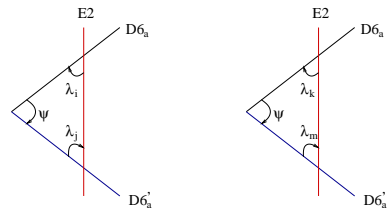}
\end{center}
\caption{\small Absorption of the zero modes}\label{su4mass}
\end{figure}

As shown in \cite{Cvetic:2007ku}, to which we refer for more details, the background of this model exhibits one class of rigid $O(1)$
instantons, whose bulk part is also shown in
Table \ref{wrapping number}.
The intersection number of these $E2$-instantons with the bulk $D6$
branes is exactly one. Taking into account the Chan-Paton label of
the gauge group $U(4)$, there are four fermionic zero modes
localised at the intersection of $E2$ and the matter branes. As
shown in figure \ref{su4mass}, these four fermionic zero modes
$\lambda$ can be saturated via two disc diagrams thereby generating
mass terms for the matter fields in the $\bf {\bar 6}$
representation.

We denote the 32 matter superfields as $\Phi^A_I=\phi^A_I+\theta \psi^A_I$,
where the lower index $I=1,\ldots, 8$  refers to the various intersections
on $T^6$ and $A=1,\ldots, 4$ counts  the different orbifold images.

As in the previous section, we can compute the disc diagrams in
figure \ref{su4mass}. Taking also the Grassmannian nature
of the fermionic zero modes into account, the overall structure of
the generated mass terms is \bea {\cal L}_{\rm mass}=  C'  M_s\,
e^{-S_{E2}}\, \,\epsilon_{ijkl}\, M^{I,J}_{A,B} \, \left(
\psi^A_I\right)_{ij}\,
    \left(\psi^B_J\right)_{kl} 
\eea
with the instanton action $S_{E2}=\frac{2 \pi }{\alpha_{SU(4)}}
\frac{V_{E2}}{V_{D6}}$. Moreover, $C'$ includes all angle dependent
constants due to the CFT-computation as well as due to
integration over all bosonic and fermionic zero modes \cite{Cvetic:2007ku,Akerblom:2007uc}. Since the
four instanton zero modes arise from the Chan-Paton factors of
$U(4)$,  the mass matrix factorizes into
\bea \label{massm}
M^{I,J}_{A,B}= h^I_A\, h^J_B,
\eea
where these factors are
essentially the disc amplitudes in figure \ref{su4mass} containing a
sum over world-sheet instantons. Due to the factorized form
(\ref{massm}), one linear combination of the 32 matter fields
receives a mass. This exemplifies that string instantons
can also generate mass terms for exotic matter fields. \\
\indent We have shown that the phenomenology of intersecting
D-brane models in Type II orientifolds requires
the inclusion of  string instanton effects.
These do generate important couplings that are
often absent perturbatively due to the strong
$U(1)$ selection rules present for D-brane models.
This becomes even more striking when the
instanton generated couplings are known
to have  certain hierarchies with respect to perturbative
couplings. Of particular interest is the appearance
of a (flipped) $SU(5)$ Yukawa coupling.
It would be very important to find globally consistent semi-realistic string vacua exhibiting this effect. \\
\emph{Acknowledgements:} We gratefully acknowledge discussions with
N. Akerblom, S. Moster, E. Plauschinn, M.
Schmidt-Sommerfeld and S. Stieberger. We also thank the
Benasque Center for Sciences for providing a nice environment for
finishing this project. This research was supported in part by the
DOE Grant DOE-EY-76-02-3071, the Fay R. and Eugene
L. Langberg Endowed Chair and the NSF Grant
INT02-03585 (M.C. and T.W.) and also by the
EU-RTN network {\sl Constituents, Fundamental Forces and Symmetries
of the Universe} (MRTN-CT-2004-005104).




\end{document}